# Modulating task outcome value to mitigate real-world procrastination via noninvasive brain stimulation

**Authors:** Zhiyi Chen [1,2,3*], Zhilin Ren [3], Wei Li [3], Zhenzhen Huo [1,2], Zhuanzheng Wang [1,2], Ye Liu [1,2], Bowen Hu [1,2], Wanting Chen [1,2], Ting Xu [1,4,5], Leonov Artemiy [6], Chenyan Zhang [7], Bernhard Hommel [7], Tingyong Feng [1,2*]

**Affiliation**

[1] Faculty of Psychology, Southwest University, Chongqing, China
[2] Key Laboratory of Cognition and Personality, Ministry of Education, China
[3] Experimental Research Center for Medical and Psychological Sciences, School of Psychology, Third Military Medical University
[4] The Clinical Hospital of the Chengdu Brain Science Institute, China
[5] Key Laboratory for Neuroinformation, University of Electronic Science and Technology of China, Chengdu, Sichuan, China
[6] School of Psychology, Clark University, Massachusetts, USA
[7] Institute for Psychological Research, Leiden University, Leiden, Netherlands

***Corresponding** at Professor Zhiyi Chen (chenzhiyi@tmmu.edu.cn) and Professor Tingyong Feng (fengty0@swu.edu.cn).

**Author Contributions:** Zhiyi Chen: Conceptualization, Methodology, Software, Writing - Original Draft and Visualization; ZhenZhen Huo, Chenyan Zhang, Bernhard Hommel: Writing - Original Draft, Formal analysis and Validation; ZhuangZheng Wang, Ye Liu, Bowen Hu, Wanting Chen: Writing - Review & Editing, Methodology, or Validation; Tingyong Feng, Conceptualization, Supervision, Project administration and Funding acquisition. CZY, RZL and LW contributed equally to this work.

**Competing Interest Statement:** All the authors disclose no competing interests.

**This PDF file includes:**

Main Text of 7,105 Words
Figures 1 to 6
Tables 1 to 2
Supporting Information (SI) Appendix 1 to 1

## Abstract
Procrastination represents one of the most prevalent behavioral problems associated with individual health and societal productivity. Despite its high prevalence and substantial impact on daily functioning, its underlying neurocognitive mechanisms remain poorly understood. A leading model posits that procrastination arises from imbalanced competing motivations: the avoidance of negative task aversiveness and the pursuit of positive task outcomes, yet this theoretical framework has not been fully validated in real-world settings and not applied effectively to guide interventions. Here, we addressed this gap with a double-blind, randomized controlled trial. We applied seven sessions of high-definition transcranial direct current stimulation (HD-tDCS) to the left dorsolateral prefrontal cortex (DLPFC) in chronic procrastinators. Using the intensive experience sampling method (iESM), we assessed the effect of anodal HD-tDCS on real-world procrastination behavior at offline after-effect (2-day interval) and long-term after-effect (6-month follow-up). We found that this neuromodulation produced a lasting reduction in real-world procrastination, with effects sustained at a 6-month follow-up. While the intervention is significantly associated with both decreased task aversiveness and increased perceived task outcome value, a mediation analysis indicated a disassociable mechanism: the increase in task outcome value (but not task aversiveness) showed a statistical pattern consistent with accounting for the observed behavioral improvement. In conclusion, these findings are consistent with the hypothesis that enhancing DLPFC function may reduce procrastination by selectively amplifying the valuation of future rewards, not by simply reducing negative feelings about the task. These results align with established decision-theoretic frameworks and suggest a targeted, theory-informed avenue for future behavioral interventions.

## Main Texts

## Introduction

Procrastination is increasingly becoming a prevalent behavioral problem around the world, which reflects the irrational voluntary postponement of scheduled tasks albeit being worse off for such delays (Blake, 2019; Steel, 2007). In epidemiological investigations, more than 15% of adults were identified as having chronic procrastination problems, and the situation for students was worse as 70-80% of undergraduates engaged in procrastination (American College Health Association, 2022; Ferrari et al., 2005). Moreover, the behavioral genetic evidence indicates a certain heritability of procrastination in human beings as well (Gustavson et al., 2017; Gustavson et al., 2014, 2015). In addition to its prevalence, the undesirable associations between procrastination behavior and health also warrant caution. There is cumulative evidence to show the close associations between procrastination behavior and work performance, financial status, interpersonal relationships, and subjective well-being (Ferrari, 1994; Pychyl & Sirois, 2016; Steel et al., 2021). Further, as prospective cohort studies indicate, many mental health problems emerge alongside procrastination, particularly sleep problems, depression, and anxiety (Hairston & Shpitalni, 2016; Johansson et al., 2023). Even worse, chronic procrastination has been consistently associated with poor general health conditions, such as immune system disruption, gastrointestinal disturbance, hypertension and cardiovascular disease (Sirois, 2015; Sirois, 2016). Thus, given these critical ramifications, considerable efforts have been devoted to delve into why we procrastinate irrationally.

To probe why we procrastinate irrationally, researchers have built upon theoretical bases of procrastination from different perspectives. For instance, Steel (2007) pioneered a promising temporal motivation theory (TMT) to explicate procrastination as the failure of self-regulation. This theory suggests that individuals would procrastinate as task utility is devalued in the far future (Steel & König, 2006). Based on insights into emotional regulation, the mood repair perspective provides another explanation to elucidate that procrastination is due to the failure of self-regulation to give priority to short-term mood repair caused by doing the task rather than long-term task reward (Sirois & Pychyl, 2013). Recently, the temporal decision model (TDM) of procrastination further provides an integrative framework to explain the procrastination decisions, which highlights that procrastination is contingent on the trade-off between task aversiveness and task-outcome value (Zhang & Feng, 2020; Zhang, Liu, et al., 2019). Task aversiveness reflects how unpleasant individuals perceive tasks to be, with more unpleasant feelings making procrastination more likely (Zhang & Feng, 2020). Task outcome value indicates how much it is worth as we evaluate the benefits it provides us (e.g., keeping body health) once we complete the task before the deadline (e.g., doing scheduled exercise) (Zhang & Feng, 2020). If the task aversiveness is overvalued in this trade-off, the decision to postpone tasks would be made consistently.

Theoretical frameworks posit that this trade-off and subsequent procrastination decisions may be modulated by individual differences in top-down regulatory capacity, with higher capacity typically associated with less procrastination (Blake, 2019; Ramzi & Saed, 2019; Zhao et al., 2021). Thus far, theoretical models have proposed three potential pathways through which regulatory processes might contribute to reduced procrastination: one for decreasing task aversiveness, the other one for increasing task-outcome value, and the last one for both (Zhang & Feng, 2020). As identified by both behavioral and neural evidence, procrastinators consistently report high task aversiveness when receiving scheduled tasks, and are more likely to postpone tasks so as to devalue negative task aversiveness (Blunt & Pychyl, 2000; Zhang et al., 2021). Meanwhile, priori literature suggests that top-down regulatory processes may facilitate the downstream modulation of negative emotional stimuli (Paschke et al., 2016; Tice & Bratslavsky, 2000). Therefore, one hypothesized pathway involves the downstream modulation of task aversiveness (Eckert et al., 2016). On the other hand, as a value-based decision, procrastination behavior is contingent on the evaluation of future outcomes (Rebetez et al., 2016; Zhang, Becker, et al., 2019). Existing evidence has shown that procrastinators generally underestimate the task-outcome value (H. Wu et al., 2016; Zhang et al., 2021). In this vein, it is hard to generate the motivation to take immediate action (Taura et al., 2015). Notably, increasing the value of future rewards has been found effective in making individuals inclined to pursue future outcomes by potentially engaging prefrontal regulatory networks to prioritize future outcomes (Cho et al., 2015; Kelley et al., 2018). Thus, another pathway worth putting forward is that procrastination behavior could be shaped by increasing future task-outcome value. Also, given the integrative nature of prefrontal regulatory functions, we hypothesize a third pathway whereby both decreased task aversiveness and increased task-outcome value may jointly contribute to reduced procrastination, potentially reflecting coordinated downstream effects on valuation and affective processing.

Consistent with this framework, the left dorsolateral prefrontal cortex (DLPFC)—a region frequently implicated in value-based decision-making and top-down regulation—has been associated with procrastination. Structural and functional variations in the left DLPFC have been correlated with procrastination tendencies (Chen et al., 2020; Hu et al., 2018; Liu & Feng, 2017). In addition to structural brain hallmarks, the neurofunctional anomalies underlying procrastination were found in DLPFC-involved circuits (Y. Wu et al., 2016; Xu et al., 2021). Furthermore, the triple brain network theory provides network-based insights to highlight the neural signature of procrastination as the self-control neural network (e.g., left DLPFC; anterior cingulate cortex, ACC), emotional regulation network (e.g., insula and orbitofrontal cortex, OFC) and episodic prospection network (e.g., hippocampus and ventromedial prefrontal cortex, vmPFC, amygdala) (Chen & Feng, 2022; Chen et al., 2020; Schlüter et al., 2018; Wypych et al., 2019). In addition to the left lateralization, there is solid evidence indicating significant associations between top-down regulatory process and the right DLPFC indeed, particularly given that this region specifically functions in top-down regulation, future self-continuity representation and social decisions (Huang et al., 2025; Knoch & Fehr, 2007; Lin & Feng, 2024). Despite this case, Xu and colleagues demonstrated null effects of anodally stimulating the right DLPFC to modulate either value evaluation or emotional regulation for changing procrastination willingness (Xu et al., 2023). Moreover, a substantial amount of neural evidence supports this conclusion that DLPFC is involved in long-term reward evaluation and value encoding via top-down regulation circuits (Frost & McNaughton, 2017; Jimura et al., 2013; Smith et al., 2018). Using a neurocomputational model, Le Bouc & Pessiglione (2022) provided clear evidence indicating that dorsal PFC signaling expected effort values was significantly attenuated in procrastinators compared to healthy controls (Le Bouc & Pessiglione, 2022). In this vein, this evidence supports this conceptualization that the left DLPFC may be a domain-specific neural signature determining one's procrastination. In light of technical advances, high-resolution transcranial direct current stimulation (HD-tDCS) has been widely used to reveal the potential neurocognitive mechanism of problematic behaviors by modulating cortical excitability, blood-brain barrier permeability and even neuroplasticity (Cirillo et al., 2017; Woods et al., 2016), which is regulated by the NMDA (N-methyl-D-aspartate) system to either bolster LTP (Long-term potentiation) or LTD (Long-term depression) processes (Chrysikou et al., 2022; Shin et al., 2020). For instance, anodic HD-tDCS applied to the left DLPFC was found effective in inhibiting problematic behaviors caused by the lack of self-regulatory ability (Allenby et al., 2018), showing significantly amplified local neural oscillations (Chrysikou et al., 2022). Beyond regional neuromodulation in a dose-free protocol, cumulative evidence has well-documented that the effects of tDCS for neuromodulation are highly dose-dependent and are involved in network-wise covariance (Sabé et al., 2024; Soleimani et al., 2023; Woodham et al., 2025). Thus, this study aims to clarify the brain-behavior association between DLPFC neuromodulation and procrastination, and to test whether observed changes in task valuation and aversiveness are consistent with theoretical models of top-down regulation.

To empirically evaluate these theoretical pathways, we conducted a double-blind, randomized, multiple-session, placebo-controlled design, with a 2 (active HD-tDCS vs. sham control) × 2 (before first neural stimulation vs. after last neural stimulation) full factorial design **(see Fig. 1)**. This HD-tDCS protocol consisted of 7 sessions spaced over 15 days, and each session was implemented every two days. To ensure sound ecological validity, each procrastinator was informed to report one REAL-LIFE task that he/she should complete the following day (e.g., Day 2) after the current neuromodulation session (e.g., Day 1), and was asked to report the ACTUAL performance for completing this task in that day (Day 2) **(see Fig. 1)**. Based on the temporal decision model of procrastination (Zhang, Liu, et al., 2019), we drew on the experience sampling method (ESM) to estimate the real dynamics of task aversiveness and task-outcome value by using a parameter-free model in each session. More importantly, to clarify which pathway best explains the neurocognitive mechanisms of procrastination, we built upon the mixed-effects general linear model and Quasi-Bayesian mediation model to test whether the changes in task aversiveness and task outcome value modulated by HD-tDCS could predict decreased procrastination. Finally, the follow-up investigation for actual procrastination was conducted for 6 months after the experiment to examine the long-term after-effect of this neural stimulation.

## Materials and Methods

This study fully adhered to CONSORT reporting guidelines and was originally preregistered in the OSF repository (10.17605/OSF.IO/Y3EDT). However, due to the technical constraint related to OSF account service **(see SI Methods)**, this OSF page is no longer accessible. For transparency and best practices of open science, based on the original protocol documentations, a preregistration statement has been reconstructed to clarify a priori hypotheses, sample size determinations, and analysis plans for this study **(Table S1)**.

### Participants

Due to the lack of diagnostic criteria for clinical procrastinators, we recruited a large-scale sample (n = 1,682) to obtain a stable benchmark distribution. Thus, the procrastinators were captured once their procrastination scores were higher than 66 on General Procrastination Scale (GPS) **(see Fig. 2A-B)**. Following this criterion, a total of 186 participants were included initially, which was in accordance with empirical evidence (i.e., 10 - 15% prevalence of procrastination) (Harriott & Ferrari, 1996). Subsequently, the semi-structured interview was performed to screen those suffering from problematic procrastination and volunteering for this study, thereby enrolling 53 participants **(see SI Methods)**. Seven participants were eventually excluded from the analyses because they voluntarily dropped out before experimental completion. All the included participants were screened for depression and anxiety symptoms **(see Tab. 1)**.

A full randomized block design was used to assign participants to both groups (active neuromodulation group, NM; sham-control group, SC) **(see Fig. 2C)**. As the pilot study probing into the effect of single-session tDCS stimulation to change procrastination willingness indicated ($t$ = 2.38, $p$ = .02, 95% CI [0.14, 1.49]; Xu et al., 2023), statistical power was predetermined by G*Power at a relatively medium effect size (1-$\beta$ err prob = 0.80, $f$ = 0.25), indicating the total sample size at 34 (17 per group) to reach acceptable power. To account for potential attrition, we determined to recruit 36 participants, at least. **(see SI Methods and Fig. S1)**. All the participants reported no history for HD-tDCS or neuromodulation. No significant differences were found between groups for any demographic characteristics **(see Tab. 1)**. This study and protocol have been fully approved by the Institutional Review Board (IRB) of the School of Psychology, Southwest University (China, IRB200301108).

### Measurement

The General Procrastination Scale (GPS) developed by Lay (1986) was used to quantify one's chronic procrastination symptom here (Lay, 1986). This scale was widely adopted in many cross-cultural contexts, and has been reported to have good psychometric properties (Klein et al., 2019). There were two additional items that we added for lie detection, including description of "the sky is red" and "I have never taken a shower". If either one was selected as "agree" by participant, his/her response would be discarded. Internal reliability of the GPS in the current study was found acceptable (Cronbach's α = 0.890). No significant difference was found between groups for GPS scores ($t$ = -1.08, 95% CI: -4.296 1.283, $p$ = .283; Jeffreys-Zellner-Siow Bayesian factor, BF, $BF_{10}$ = 0.455, error = 0.020 %).

### Experimental design and procedure

#### Nested cross-sectional longitudinal design

This study used a nested cross-sectional longitudinal design to investigate whether the multiple-session anodal HD-tDCS targeting the left DLPFC could reduce actual procrastination behavior and to probe how this effect manifests. To assess procrastination in daily life, we implemented a 15-day protocol alternating between Neuromodulation Days (Days 2, 4, 6, 8, 10, 12, 14) and Task Days (Days 1, 3, 5, 7, 9, 11, 13, 15). On the Neuromodulation days, the 20-min anodal HD-tDCS neuromodulation targeting the left DLPFC was performed for the HD-tDCS active group at intervals of 2 days, while the sham-control group received sham HD-tDCS training. This HD-tDCS training was repeated for a total of seven sessions, and lasted 15 days **(see Fig. 1)**. Crucially, to capture procrastination in ecologically valid contexts, priori to receiving either active or sham HD-tDCS (administered between 09:00–18:00), participants were instructed to specify a real-life task they were personally obligated to complete the following day, with a self-defined deadline strictly constrained to 18:00–24:00 to ensure ≥24 hours between stimulation offset and task deadline, thereby isolating offline after-effects. This task should meet the following three criteria: (a) it should be already assigned in the real-world settings; (b) deadline should be constrained to 18:00-24:00 (see above); (c) it should be more likely to induce procrastinate. By doing so, more than 300 real-life tasks were collected, spanning academic (e.g., "submit a statistics homework assignment"), occupational (e.g., "draft and email a project proposal"), administrative (e.g., "complete online application for Class C driver's license"), self-improvement (e.g., "practice guitar for ⩾30 minutes"), domestic (e.g., "do laundry"), and health-related (e.g., "running 2,000m for exercise"). Full task list has been tabulated in the Appendix 1. As primary outcomes, all the participants were required to report task-execution willingness (TEW) (Zhang & Feng, 2020; Zhang, Liu, et al., 2019), for a real-life task 24 hours post-neuromodulation. Thus, procrastination willingness was quantified as 100-TEW score (see underneath for details). Furthermore, we asked participants to report the actual task completion rate (CR) of the task at the deadline (e.g. participant A finished 90% homework at deadline and reported this situation to us at deadline). In this vein, the actual procrastination rate (PR) was quantified as 1-CR.

On the Task Day, we developed a mobile app to implement experience sampling method (ESM) for tracking one's real-time evaluation of task aversiveness and task outcome value (**see Fig. 1**). The task aversiveness describes how disagreeable one perceives performing a given real-life task to be, whereas outcome value refers to the subjective benefits of the task outcome brought about by completing the task before the deadline (Zhang & Feng, 2020). As theoretically conceptualized by the temporal decision model (TDM) of procrastination, the perceived task aversiveness is hyperbolically discounted when approaching deadline, showing sharply discounting when faring away from deadline but slowly discounting once nearing deadline (Zhang & Feng, 2020). Thus, considering this nonlinear dynamics inherent in this hyperbolic discounting, the five recording moments of ESM were selected per task a priori by using a log-spaced temporal sampling scheme (Myerson et al., 2001), with increasing sampling density toward the deadline, such as moments of 10:00 (earliest), 16:00, 18:00, 19:30, 20:00 (deadline). The five sampling points could meet statistical prerequisite in the hyperbolic model fitting, requiring ≥ 4 points (Green & Myerson, 2004). To do so, recording moments of tasks were individually tailored for each task per participant in this ESM procedure. To obviate the confounds of daily emotions in task aversiveness evaluation, we used the averaged scores of PANAS at 10:00 (noon) and 16:00 (afternoon) as anchoring points to quantify one's daily emotions by using this ESM app. Before each session of HD-tDCS training, each participant was required to report a real-life task whose deadline is tomorrow. To obtain the long-term effect of HD-tDCS (i.e., the interval between HD-tDCS and task completion is at least 24 hours), the task deadline that participants reported was required to be between 18:00 - 24:00. Once a sampling time reached, this app would send a digital message to require participants to fill online form for data collection.

**Quantification of covariates of interests**

Outcome variables of this study were twofold: one is task-execution willingness and another is procrastination rate (PR). Task-execution willingness is used to evaluate one's subjective inclination to avoid procrastination (Zhang & Feng, 2020). In this vein, we used a 100-point scale to require participants to report their task-execution willingness (0 for "I will definitely procrastinate this task" and 100 for "I will take action to complete this task immediately"). This metric was recorded 24 hours after neuromodulation to examine its long-term effects. PR is used to quantify the extent to which one task has been procrastinated, and was calculated as 1 - CR (task completion rate). Critically, at the precise deadline, the app prompted participants to (a) indicate task completion status (yes/no), and if incomplete, (b) report the percentage completed (1–99%), defined as the Task CR, while simultaneously uploading objective evidence (e.g., screenshots of submitted files, photos of physical outputs, system-generated logs, or app-exported records). If the task was actually completed before the deadline, the CR would be 100% and the PR would be calculated as 0% (1-CR). PR was recorded at the actual task deadline for each participant. We were also interested in re-investigating their actual procrastination by using PR 6 months after the last neuromodulation to test the long-term after-effect of this neuromodulation.

From what has been mentioned above, task aversiveness and outcome value were considered key factors to explain the effect of neuromodulation on reducing procrastination in the current study. To quantify one's task aversiveness, participants were required to rate their feelings towards the task by using a 100-point visual analog scale (i.e., How do you feel in the current moment when you need to complete this task before deadline, with 0 for "extremely unpleasant", 50 for "totally neutral" and 100 for "extremely pleasant"). Likewise, participants are also required to rate their outcome value by using a 100-point visual analog scale (i.e., How much do you desire to obtain incentive outcome of this task, with 0 for "extremely weak", 50 for "medium" and 100 for "extremely strong"). As articulated temporal decision theoretical model above, the task aversiveness evoked by executing a task was temporally dynamic in a hyperbolic discounting pattern, with sharply discounting in faring away from deadline but slowly discounting in nearing deadline (Zhang & Feng, 2020). To quantitatively characterize the task aversiveness with consideration for its dynamics, the model-free area under the curve (AUC) was calculated. Specifically, based on the log-spaced temporal sampling rule, task aversiveness was measured by 100-point visual analog scale at the five sampling moments. Then, the task aversiveness discounting (A) was calculated as 1- (A(t) / A(earliest)), where t(earliest) was the earliest sampling point, serving as the reference for immediate execution. Subsequently, using the GraphPad Prisma software (v9, 525), the AUC was computed as the trapezoidal integration between task aversiveness discounting and time across five data points, basing on the Myerson algorithm (Myerson et al., 2001). By doing so, a higher AUC reflects stronger temporal discounting of task aversiveness along with nearing deadline, which means that participants experience a faster decline in subjective aversiveness as execution is delayed, yielding lower effective aversiveness and reduced avoidance behavior. As for the task outcome value, it was theoretically posited as a relatively stable evaluation of the task (Zhang & Feng, 2020). Therefore, it was quantified by the self-reported 100-point visual analog scale after neuromodulation at least 12 hours later, to ensure no online effect.

**HD-tDCS protocol**

The HD-tDCS suit (stimulater and 4 × 1 multichannel stimulation adapter, MSA) that this study used was produced by the Soterix Medical Inc., and has been widely verified safe, effective and reliable for public (Villamar et al., 2013). Based on advanced properties of 4 × 1 MSA, the targeted areas for current flow can be constrained within 2.5 $cm^2$ (Villamar et al., 2013).

To position electrodes into targeted areas (left DLPFC), the 10-20 international system for EEG was initially used to mark potential nodes, and determined Cz as reference point. There is compelling evidence to claim that the left F3 could be used as target node for modulating the left DLPFC (Seibt et al., 2015; Tsukuda et al., 2025). In this vein, the central anodal electrode was determined onto F3, and four return electrodes surrounded central electrode at outside of 7.5 cm, including F5, AF3, FC3 and F1 **(see Fig. 2E)**. Ramp-up and ramp-down durations were set to 30 seconds. To further locate the targeted areas, the high-definition neuronavigation system (ANT Neuro Inc., Welbergweg, Germany) was performed. Results indicated the accurate position that we pre-determined by showing a high overlap probability over the left DLPFC (MNI Coordinate: -51 40 18, 94.33 % overlapping probability) **(see SI Method and Fig. 2E)**. In addition, the coordinates of this targeted area were retrieved from the Brede Database (http://neuro.imm.dtu.dk/services/), and showed highly pertinent functions related to DLPFC.

Before stimulation, participants were informed to clean the scalp to reduce resistance. Then, cotton swabs were used to separate hair until the scalp surface become visible. Subsequently, the electrically conductive gel (about 1.5 ml) was introduced into the plastic casing facilitating constraint of current flow. Next, the Ag/AgCI sintered ring electrodes were placed onto plastic casing and covered with a cap to lock them in the right positions. To reduce discomfort, the electrode cables were taped elsewhere. Further, the above processes would be re-adjusted if any electrode resistances were found larger than 1.5 units (Villamar et al., 2013). Once all the processes had been completed, the stimulator would be launched.

Participants in the HD-tDCS training group underwent constant electric current of 2.0 mA targeting the left DLPFC for 20 minutes. Results from the simulation of electric density showed a peak current of ~0.5mA/cm $^2$ at the central electrode and of ~0.125 mA/cm$^2$ at the four return electrodes, thereby indicating the safety and effectiveness. As for the sham-controlled group, the stimulator would deliver current flow with 2.0 mA during the first and last 30 seconds to elicit sense of electric stimulation for blinding of them. To obtain the pure offline effect, these measures for task-execution willingness, task aversiveness, and outcome value were conducted after stimulation at least 12 hours (Bikson et al., 2016).

**Statistics**
All the statistics were implemented by *R* (https://www.rstudio.com/) and *R*-dependent packages.

To clarify whether multiple-session HD-tDCS neuromodulation can reduce procrastination, the general linear mixed-effects model (LMM) was constructed for subjective procrastination willingness (i.e., self-reported visual analog scores) and actual procrastination behavior (i.e., real-world task-completion rate before deadline). Here, sex, age and socioeconomic status (SES) were modeled as covariates of no interest. As the National Bureau of Statistics (China) issued (https://www.stats.gov.cn/sj/tjbz/gjtjbz/), on the basis of per capita annual household income, the SES was divided into seven hierarchical tiers from 1 (poor) to 7 (rich). To obviate subjective rating bias stemming from individual daily mood, we separately measured participants' daily emotional fluctuation at 10:00 and 16:00 using a self-rating visual analog item (i.e., "How do you feel today?", 0 for "completely uncomfortable" and 100 for "definitely happy"). By doing so, the averaged score of those self-rating emotions at the two time points was modeled into the LMM as a covariate of no interest. Given the risks of convergence failure with the two correlated random-effects structure (i.e., treatment days and self-reported emotions), we hypothesized that the random effects are independent, leading to model simplification. Consistent with this assumption, model comparisons favored the simplified independent structure over the full correlated structure for both outcomes. For the actual procrastination model, the simplified model showed lower AIC ($\Delta$ = -1.0) and BIC ($\Delta$ = -11.8), with a non-significant likelihood ratio test ($\chi^2(3) = 5.04$, $p = .17$). For the task-execution willingness model, the simplified model was also preferred ($\Delta$AIC = -5.5, $\Delta$BIC = -16.4; LRT: $\chi^2(3) = 0.51$, $p = .91$). This analysis was implemented using the "lme4" and "lmerTest" packages. Employing "emmeans" package, simple effects were also tested at baseline and post-last-intervention using Tukey-adjusted pairwise comparisons of estimated marginal means from the full LMM, controlling for covariates and random-effects structure. To validate statistical robustness, instead of continuous outcomes for parametric tests, we also conducted a between-group comparison for the number of tasks that procrastination emerges by using the nonparametric $x^2$ test with φ correction or Fisher exact test. Regarding the 6-month follow-up investigation, this LMM was also built to examine the long-term after-effect of neuromodulation on reducing actual procrastination. To examine whether our findings were sensitive to the distributional assumptions of the dependent variables, we re-analyzed the main models using an alternative distributional specification. Given that both procrastination rates (ranging from 0% to 100%) and task-execution willingness (measured on a 0-100 visual analog scale) are bounded continuous outcomes, and that procrastination rates frequently reached the lower boundary (0%) in the present study, a Beta regression model with a logit link function was employed for a sensitivity analysis.

To ascertain the neurocognitive mechanism of tDCS in reducing procrastination, the Quasi-Bayesian mediation analysis was used to model the association between the effects of tDCS, task aversiveness/outcome and decreased procrastination. To build upon this model, the tDCS treatments were inputted as independent variables, and the task aversiveness/outcomes were modeled as mediating variables by using the "Mediation" package (https://cran.r-project.org/web/packages/mediation/) (Imai et al., 2010). We estimated these pathway effects (i.e., average mediation effects, δ; average direct effect, ζ; total effects, ρ) by using Markov Chain Monte Carlo (MCMC) sampling. To improve the statistical reliability, the sequential ignorability assumption was tested by using sensitivity analysis. Details for the statistical principals and basis could be found elsewhere (Imai et al., 2010). As these mediation analyses are based on observational measures rather than experimentally manipulated mediators, they do not establish causal pathways. Simulation studies have shown that such analyses can suffer from inflated Type 1 error rates. Results should therefore be interpreted as hypothesis-generating and statistically consistent with the proposed theoretical model, rather than as confirmatory evidence of causal mechanisms.

## Results

### Blinding

In both groups, almost all participants reported perceiving acceptable pain stemming from current stimulation, and believed they were receiving treatment, with 91.30% (21/23) for active neuromodulation group (NM) and with 86.95% (20/23) for sham control group (SC) ($x^2$ = 0.224, $p$ = .636). All the participants were engaged in the identical experimental procedures excepting stimulation's type (active *vs* sham). In addition, statistical models excluded Session 1 and Session 4 because participants reported additional unexpected events that uncontrollably disrupt task execution in both groups **(see SI Result and Tab. S2)**.

### Multiple-sessions HD-tDCS (ms-tDCS) can alleviate procrastination

To identify whether ms-tDCS targeting the left DLPFC can alleviate subjective procrastination willingness and actual procrastination behavior, a general mixed-effects linear model (LMM) with Satterthwaite's method was built, with task-execution willingness and actual procrastination rates (PR) as primary outcomes, respectively. For procrastination willingness, results showed a statistically significant interaction effect between multi-session neuromodulations and groups ($\beta$ = -7.84, SE = 1.80, $t$ = -4.36, DF = 45.6, $p$ < .001, Cohen $d$ = 2.37, 95% CI: 1.49-3.25; **Fig. 3A and Fig. S2a**). In the post-hoc simple effect analysis, it demonstrated a significantly increased task-execution willingness (i.e., decreased procrastination willingness) after neuromodulation in the active neuromodulation group (NM-before: 35.65 ± 30.21, NM-after: 80.43 ± 19.92, Mean Diff = 41.79, SE = 7.58, DF = 103.4, *t*.ratio = 5.51, $p$ < .0001, Tukey correction), but no such effects were identified in the sham control group (SC-before: 37.57 ± 26.46, SC-after: 47.35 ± 30.49, Mean Diff = 2.58, SE = 7.56, DF = 96.8, *t*.ratio = 0.34, $p$ = .73, Tukey correction) **(Fig. 3B-C)**. A linear uptrend for task-execution willingness was further observed across multiple sessions in the active NM group, indicating gradually increasing neuromodulation effects **(Fig. 3D;** $p$ < .01, Mann-Kendall test**)**. For actual procrastination behavior, changes to actual procrastination rates across all the sessions have been detailed in the **Fig. 3E and Fig. S2b.** Similarly, a statistically significant interaction effect was identified here ($\beta$ = -7.37, SE = 2.40, $t$ = -3.02, DF = 46.6, $p$ = .004, Cohen $d$ = 1.52, 95% CI: 0.87-2.16), and the simple effect analysis further revealed decreased actual procrastination rates after ms-tDCS in the active neuromodulation group (NM-before: 56.74 ± 39.10, NM-after: 0.00 ± 0.00, Mean Diff = 44.40, SE = 9.36, DF = 110.0, *t*.ratio = 4.74, $p$ < .0001, Tukey correction), but no such prominent changes found in the sham control group (SC-before: 46.47 ± 40.76, SC-after: 33.35 ± 37.82, Mean Diff = 7.53, SE = 9.28, DF = 102.0, *t*.ratio = 0.81, $p$ = .42, Tukey correction) **(Fig. 3F-G)**. Also, a significant downtrend for procrastination rates across all the sessions was identified in the active NM group **(Fig. 3H;** $p$ < .01, Mann-Kendall test**)**.

Furthermore, the nonparametric $x^2$ test of R × C contingency table was conducted for the count of procrastinated tasks, by treating this outcome (i.e., whether a participant actually procrastinates the task in real-world settings) as an ordinal variable. Results showed a significant reduction in group-averaged procrastination frequency in the active neuromodulation group after last-session HD-tDCS, but not in the sham control group (NM-before: 69.56% (16/23 participants), NM-after: 0.00% (0/23 participants); SC-before: 69.56% (16/23 participants), SC-after: 56.52% (13/23 participants), $x^2 = 10.08$, $p < .001$), partly indicating a high statistical robustness. To systematically test whether such effects are biased by extreme data points or patterns, we reran the main LMMs by iteratively removing data from the last two sessions, which showed extraordinarily high effectiveness from neuromodulation (e.g., all the participants in the neuromodulation group had no actual procrastination behavior in session #6 and #7). Results showed the significant group*neuromodulation sessions interaction effects across all those nested models (removing session #6, #7 or both, all $p < .05$; **see SI Results and Tab. S3-4**), potentially indicating a statistical robustness from the data pattern. Furthermore, as a sensitivity analysis, we reran the main LMMs using Beta regression distribution with a logit link function, which is appropriate for both bounded outcomes mentioned above (i.e., procrastination rate and procrastination willingness). The main effects (i.e., Group and Treatment day) and their interaction remained significant for both procrastination rate and willingness **(see SI Results and Tab. S5)**, confirming that our findings are robust to alternative distributional assumptions. In brief, these findings provided empirical evidence to support that ms-tDCS neuromodulation targeting the left DLPFC can be an effective way to reduce both procrastination willingness and actual procrastination.

**Ms-tDCS changes task aversiveness and task-outcome value**

Both task aversiveness and task outcome value serve as key pathways determining whether one would procrastinate. To this end, we further utilized a generalized linear mixed-effects model to examine the effects of ms-tDCS on changes in task aversiveness and task outcome value. Task aversiveness changes across all the sessions are shown in the **Fig. 4A/C and Fig. S3**. We demonstrated a statistically significant decrease in task aversiveness and an increase in task outcome value via ms-tDCS in the neuromodulation group (Task aversiveness: interaction effect, $\beta = -0.12$, SE = 0.04, DF = 46.69, $t = -3.28$, $p = .002$; simple effect, NM-before $_{(AUC)}$: 1.13 ± 0.54, NM-after $_{(AUC)}$: 1.95 ± 0.84, Mean Diff = 0.81, SE = 0.16, DF = 104.5, *t*.ratio = 4.91, $p < .001$, Tukey correction; Outcome value: $\beta = -6.76$, SE = 1.74, DF = 46.2, $t = -3.89$, $p < .001$; simple effect, NM-before: 35.87 ± 27.83, NM-after: 73.09 ± 23.34, Mean Diff = 39.36, SE = 7.25, DF = 103.7, *t*.ratio = 5.43, $p < .001$, Tukey correction; **see Fig. 4B**), but not in the sham control group (Task aversiveness: SC-before $_{(AUC)}$: 1.07 ± 0.51, SC-after $_{(AUC)}$: 1.28 ± 0.46, Mean Diff = 0.17, SE = 0.17, DF = 98.3, *t*.ratio = 1.06, $p = .29$, Tukey correction; Outcome value: SC-before: 34.00 ± 25.17, SC-after: 40.13 ± 28.94, Mean Diff = 5.54, SE = 7.26, DF = 98.7, *t*.ratio = 0.76, $p = .44$, Tukey correction; **see Fig. 4D**). In the neuromodulation (NM) group, task aversiveness steadily decreased with the cumulative number of stimulation sessions, while perceived task outcome value increased significantly **(see Fig. 4E-F**, $p < .05$, Mann-Kendall test). Thus, these findings are consistent with the view that neuromodulation of the left DLPFC is associated with reduced task aversiveness and increased task-outcome value.

**Increased task outcome value but not decreased task aversiveness predicts reduced procrastination**

Given the dual neurocognitive pathways identified above—reduced task aversiveness and increased task-outcome value—we proposed that these changes may reflect downstream consequences of prefrontal neuromodulation on value-based decision processes, statistically consistent with theoretical models of top-down regulation. To this end, we utilized a general linear model to regress decreased task aversiveness and increased task outcome value to changes in task-execution willingness. In this model, increased task outcome value ($\Delta_{\text{Outcome value}}$) significantly predicted increased task-execution willingness ($\Delta_{\text{task-execution willingness}}$) ($x^2 = 15.95$, $p < .01$, $R^2 = .40$; $\Delta_{\text{Outcome value}}$: $\beta = 0.61$, S.E. = 0.12, $p < .001$, 95% CI: 0.37 - 0.86), whereas no significant effect was observed for predicting task-execution willingness through decreased task aversiveness ($\Delta_{\text{Task aversiveness}}$: $\beta = 0.10$, S.E. = 0.12, $p = .41$, 95% CI: -0.14 - 0.34). Likewise, for actual procrastination behavior in real-world settings, increased outcome value ($\Delta_{\text{Outcome value}}$) was identified to be significantly predictive, whereas decreased task aversiveness showed null effects **(see Tab. 2)**. Collectively, these findings provide statistical evidence consistent with the hypothesis that the outcome-value pathway may contribute to procrastination reduction.

**Increased task outcome value is specifically associated with reduced procrastination in the contexts of neuromodulation**

To explore the potential neurocognitive pathways of procrastination, the Quasi-Bayesian mediation analysis was undertaken, with increased task outcome value as a statistically mediated variable. As an exploratory analysis, results indicated that increased task outcome value was associated with changes in the task-execution willingness ($\delta$ = 21.73, p < .01; $\zeta$ = 11.25, p = .07, $\rho$ = 32.99, p < .01, simulation = 1,000; **see Fig. 5A)** and real-world procrastination ($\delta$ = 30.75, p < .01; $\zeta$ = 3.05, p = .52, $\rho$ = 33.81, *p* < .01, simulation = 1,000; **see Fig. 5B)**, in the context of ms-tDCS neuromodulation. To ensure the statistical robustness and specificity of these findings, the sensitivity analysis was implemented by changing sampling parameters and outcome variables. By doing so, those findings were validated statistically robust, as shown by replicated observations across bootstrapping sampling subsets **(see SI Results and Tab. S6-7)**. Moreover, the results of the control analysis further validated the specificity of these findings by showing a null statistically mediated effect of this model to predict one's task aversiveness **(see SI Results and Tab. S8)**. In summary, these findings identified a statistical pathway consistent with the theoretical model: neuromodulation of the left DLPFC was associated with increased task-outcome value, which in turn was associated with reduced procrastination. Nevertheless, all mediation findings are now explicitly labeled as "exploratory" and framed as quantifying statistical associations consistent with the TDM pathway, not causal mediation.

**Long-term effects of ms-tDCS**

We have also attempted to conduct a follow-up investigation to test the long-term after-effect of ms-tDCS in reducing actual procrastination. Almost all the participants had undergone follow-up except one in the neuromodulation group after last neuromodulation for 6 months ($N_{NM}$ = 22, $N_{SC}$ = 23). Thus, the LMM was constructed, with the PR before first neuromodulation vs. PR after last neuromodulation for 6 months as covariates of interest. Results showed the statistically significant group*time interaction effects ($\beta$ = 16.5, SE = 9.9, *p* = .049). Simple-effect model demonstrated a decrease in actual procrastination rates in the active neuromodulation group after last stimulation for 6 months compared to baseline ($\beta$ = -22.05, SE = 10.0, *p* = .038, Tukey correction; NM-before: 40.68 ± 37.96, NM-after$_{6\text{-months}}$: 18.63 ± 29.80), and revealed null effects in the SC group ($\beta$ = 1.26, SE = 9.78, *p* = .99, Tukey correction; SC-before: 46.47 ± 40.75, SC-after$_{6\text{-months}}$: 47.73 ± 39.18) **(see Fig. 6)**. Furthermore, using a nonparametric $x^2$ test to compare differences in the number of procrastinated tasks, we still found a statistically significant reduction in procrastination frequency in NM group after neuromodulation for 6 months compared to baseline ($x^2$ = 3.30, *p* = .035, NM-before: 68.19% (15/22), NM-after$_{6\text{-months}}$: 40.91% (9/22)), while no significant changes were observed in the SC group ($x^2$ = 0.11, *p* = .74, SC-before: 69.56% (16/23), SC-after$_{6\text{-months}}$: 73.91% (17/23)). Therefore, beyond short-term effects, the benefits of ms-tDCS neuromodulation on reducing procrastination were still detectable at a 6-month follow-up, providing preliminary evidence consistent with long-term after-effects.

## Discussion

In the current study, by performing anodal ms-tDCS neuromodulation on the left DLPFC, both procrastination willingness and actual procrastination behavior were significantly decreased in real life. Additionally, a 6-month follow-up investigation revealed the long-term retention of such effects. Furthermore, this neuromodulation was found to decrease task aversiveness and increase outcome values; notably, only increased task-outcome value could predict decreased procrastination. On balance, our findings provide evidence consistent with the hypothesis that neuromodulation of the left DLPFC is associated with reduced procrastination, primarily through increasing task-outcome value rather than merely reducing task aversiveness. In addition, this study provided an effective way to reduce actual procrastination by using ms-tDCS neuromodulation.

One contribution of this study is to provide empirical evidence partially consistent with the temporal decision model (TDM), showing that increased task-outcome value—rather than decreased task aversiveness—was statistically associated with reduced procrastination following DLPFC neuromodulation. Neurobiological substrates of procrastination have been investigated in recent years, and have demonstrated the crucial roles of the left DLPFC in predicting procrastination (Chen & Feng, 2022; Chen et al., 2021; Chen et al., 2022; Hu et al., 2018; Liu & Feng, 2017; Zhang et al., 2017; Zhang et al., 2016). Not only the brain functional anomalies of the left DLPFC but also the neuroanatomical disruptions of self-regulatory brain network constituted by the left DLPFC were found to be linked with more procrastination behaviors (Chen & Feng, 2022; Zhang et al., 2016). Notwithstanding this, it still remains unclear to claim their brain-behavior relationship - that is - no known evidence existed to clarity whether changes of the left DLPFC lead to procrastination or vice versa. The current study demonstrated the role of the left DLPFC in procrastination by showing that the neuromodulation of the left DLPFC indeed manipulated procrastination, and thus provided straightforward and powerful evidence to fill this gap.

It has long been acknowledged that the left DLPFC is consistently implicated in top-down regulatory processes and value-based decision-making, such as patience to wait long-term gratification for a delay, inhibition of impulsiveness, and control of game addiction (Cohen & Lieberman, 2010; Lin & Feng, 2024). Furthermore, the increased activation of the left DLPFC has been observed during the exertion of self-regulation which modulates value signals (Hare et al., 2009; Harris et al., 2013). Meanwhile, procrastination has been argued to be the consequence of self-regulation failure for a long time (Ariely & Wertenbroch, 2002; Rebetez et al., 2018; Rozental & Carlbring, 2014). Supporting this, both the brain morphological disruptions in the DLPFC and anomalies in the functional coupling of the DLPFC-based regulatory network have been correlated with procrastination tendencies (Xu et al., 2021; Yang et al., 2021). Moreover, it was worth noting that the increased activation of the left DLPFC was found to be involved in outcome value evaluation through self-control regulation (Chen et al., 2018; Zha et al., 2019). There is more straightforward evidence to substantiate the role of manipulating the DLPFC in changing one’s subjective value evaluation (Huang et al., 2017; Minati et al., 2012). Moreover, the theoretical explanations and empirical evidence have increasingly converged into one line for claiming that the increased task outcome value would prompt more motivation to drive one to take action immediately and thus reduce procrastination (Zhang et al., 2019).  Building on this foundation, among several theoretical interpretations and cognitive pathways, our study showed the one plausible neurocognitive mechanism of procrastination: the cortical excitability of the DLPFC produced by active neuromodulation may engage prefrontal regulatory networks to increase task outcome value, which in turn is associated with reduced procrastination behavior, statistically supporting the theoretical accounts of temporal decision model (TDM, Zhang et al., 2019). TDM uniquely posits procrastination as contingent on the trade-off between task aversiveness and task-outcome value; our observation that only increased outcome value—not decreased aversiveness—statistically predicted reduced procrastination aligns precisely with TDM’s hypothesis that value-based processes may dominate affective-avoidance processes in driving behavioral change. However, neither TMT (which emphasizes temporal discounting of utility per se) nor the mood repair perspective (which prioritizes short-term affect regulation) explicitly predicts this dissociable pattern. Despite statistically supporting the TDM, we acknowledge that alternative neurocognitive mechanisms could contribute to the observed reductions in procrastination. For instance, repeated exposure to the experience-sampling protocol may have enhanced participants’ awareness of task progress or facilitated feedback-based learning, thereby increasing the subjective value of goal completion independent of DLPFC neuromodulation. Participants in the active group may have learned during the repeated sessions that completing tasks feels internally rewarding, thereby benefiting from a form of “learned industriousness” (Eisenberger, 1992) that persists months later. Similarly, improvements in working memory for task maintenance, attentional allocation to reported goals, or strategic shifts in goal selection across sessions could plausibly mediate the intervention effects. While our double-blind, sham-controlled design and inclusion of daily emotional covariates help mitigate some non-specific confounds, the present study did not incorporate direct measures of these alternative processes. Consequently, we cannot definitively isolate the value-based pathway posited by the TDM from concurrent contributions of attention, learning, or executive functions.

In addition, another contribution of the current study is to provide an effective way to reduce both procrastination willingness and actual procrastination in real-life tasks. As mentioned above, despite the fact that multifarious behavioral interventions and evidence have been massively studied for overcoming procrastination, they have shared a common aim - that is - reducing the intention-action gap (Miao et al., 2024; van Hooft et al., 2005). Eerde and Klingsieck (2018) put forward an insightful standpoint established by meta-analytic evidence: procrastination is characterized as an intention-action gap rather than an intention to postpone (van Eerde & Klingsieck, 2018). This notion has been partly supported by showing that cognitive behavior therapy (CBT) for goal-directed behaviors may outperform other interventions focusing on time management (Rozental et al., 2018). Moreover, the trans-theoretical model of procrastination has shown that the behavioral intervention may be effective in changing one’s motivation to overcome procrastination but not in actual behaviors (Grunschel & Schopenhauer, 2015). Thus, the ms-tDCS protocol employed here may offer a promising avenue for reducing procrastination by attenuating the intention-action gap, though further validation in larger, clinically screened cohorts is warranted. Furthermore, both 2-day-interval long-term effects and the 6-month long-term retention of the effects of ms-tDCS on reducing actual procrastination have been revealed as well. Thus far, the trends in adopting tDCS neuromodulation techniques in many aspects of behavioral therapies have emerged, but concern for a long-lasting effect of single session stimulation has continued (Brunoni et al., 2013; Brunoni et al., 2012). To tackle this concern well, instead of single-session tDCS, the current study adopted multiple-session stimulation to implement neuromodulation on the left DLPFC, which facilitates long-term effects (Au et al., 2017; Tedesco Triccas et al., 2016). Existing neurobiological theories and empirical evidence have demonstrated that multiple-session tDCS stimulation could boost cumulative effects of consolidation for activity-dependent LTP (long-term potentiation), which is crucial to neurobehavioral learning, and thus produce robust long-term after-effects (Agboada et al., 2020; Au et al., 2017). Intriguingly, the activity-dependent LTP process produced by multiple-session consolidation was found to contribute to long-term cortical plasticity, especially in the DLPFC (Jannati et al., 2023; Siebner & Rothwell, 2003). Thus, the detectable effects at 6 months are consistent with the hypothesis that repeated neuromodulation may induce neuroplastic changes in the DLPFC that support sustained behavioral change. However, we explicitly note that a single follow-up timepoint cannot establish the stability or trajectory of these effects; future studies with multiple longitudinal assessments are required to substantiate claims about long-term retention. On balance, this study provided an effective way to help procrastinators overcome actual procrastination in real-life.

**Limitations**

While the use of a multi-session design and the long-term assessment can be considered a strength of the present study, it also has several limitations. Even though various tDCS effects have been demonstrated so far, they also tend to be difficult to replicate and sensitive to not yet fully understood context conditions and interindividual differences, which also applies to transcranial magnetic stimulation (TMS) (Valle et al., 2009). To overcome this shortcoming, it will be necessary to establish an individually tailored tDCS protocol to improve the sensitivity of corresponding interventions (Chew et al., 2015). Thus, future research could further improve the effects of tDCS on reducing procrastination by adopting more individualized tDCS protocols. Another limitation is the lack of real-time functional neuroimaging measures to better monitor the impact of our intervention. In the absence of such measures, we had to rely on behavioral indicators to assess the success of the tDCS training. In addition to technical limitations, a major limitation of the current study is the relatively small sample size (total $N$ = 46). While this was determined a priori based on our specific pilot study, small samples inherently bear a higher risk of being influenced by outliers and may overestimate effect sizes compared to large-scale meta-analytic expectations for tDCS. Therefore, these findings warrant caution in generalization and necessitate rigorous replication in larger, adequately powered cohorts. Also, considering the lack of medical screening for psychiatric conditions (e.g., ADHD or depression) in this sample, it remains unclear whether these training effects are domain-specific for procrastination. Notably, we explicitly acknowledge that exploratory

Quasi-Bayesian mediation analyses, based on observational measures rather than experimentally manipulated mediators, cannot establish causal pathways and are susceptible to inflated Type 1 error rates as demonstrated in recent simulation studies. These findings should be interpreted strictly as hypothesis-generating and statistically consistent with the proposed theoretical model, rather than as confirmatory evidence of causal mechanisms. Substantiating the precise neurocognitive pathway will require future studies employing stronger causal designs, such as experimental manipulation of task valuation or longitudinal cross-lagged modeling. Moreover, this study did not collect data for assessing participants' self-control at either baseline or post-neuromodulation. Accordingly, we explicitly note that self-control was not directly measured or manipulated in this study; the observed associations between DLPFC neuromodulation, task-outcome value, and procrastination are consistent with theoretical models positing a role for top-down regulatory processes, but do not constitute direct evidence that self-control mechanisms were engaged. This limitation precludes definitive conclusions about the unique contribution of self-control-related pathways versus alternative neurocognitive mechanisms. In addition, despite instructing to report valid real-life tasks with high probabilities to procrastinate, we did not measure the task difficulty and consistency across sessions for each participant. Consequently, interpreting the effects of neuromodulation to mitigate procrastination as "unique contributions" warrants caution. Given the absence of cathodal HD-tDCS stimulation as a contrast condition in inference, it also warrants caution that the increased DLPFC excitability may not be the exclusive neural mechanism for procrastination. Finally, we explicitly note that a single 6-month follow-up timepoint cannot definitively establish the stability or trajectory of these effects. Future studies incorporating multiple longitudinal assessments (e.g., 1-month, 3-month, 6-month, 12-month) are required to substantiate claims about long-term retention and to disentangle the precise contributions of neuroplasticity versus behavioral learning.

## Conclusions

In conclusion, this study potentially provides an effective way to reduce both procrastination willingness and actual procrastination behavior by using neuromodulation on the left DLPFC. Furthermore, such effects have been observed for 2-day-interval long-term after-effects, and were also found for 6-month long-term retention in part. More importantly, this study identified that the ms-tDCS neuromodulation could decrease task aversiveness and increase task outcome value, and further demonstrated that the increased task outcome value could predict decreased procrastination, a relationship that is conceptually consistent with theoretical models of top-down regulation and value-based decision-making. In this vein, the current study enriches our understanding of neurocognitive mechanism of procrastination by showing the prominent role of increased task outcome value in reducing procrastination. Also, it may inform the development of theory-driven, neuromodulation-informed strategies for behavioral interventions, pending further validation in diverse populations.

## Acknowledgement

We appreciate Zhibing Xiao (Beijing Normal University) for assistance with programming and coding; Chao Ran (Alibaba technical team) for assistance with developing experience sampling online platform; YanCheng Tang (Peking University), Dr. Peikai Li (Utrecht University) and Dr. Hu Chuan-Peng (Nanjing Normal University) for assistance with statistics.

## Funding

This study was supported by the National Natural Science Foundation of China (32300907; 32571253, 32271123), Key Projects for Technological Innovation and Application Development in Chongqing [CSTB2022TIAD-KPX0150], the National Key Research and Development Program of China [2022YFC2705201], and Innovation Research 2035 Pilot Plan of Southwest University [SWUPilotPlan006].

## Data and Code Availability

We report how we determined our sample size, all data exclusions (if any), all manipulations, and all measures in the study, and the study follows JARS (Appelbaum, et al., 2018). Data, analysis code, and research materials are available at Science Data Bank (ScienceDB, https://doi.org/10.57760/sciencedb.35140). Data were analyzed using R, version 4.4.1 (R

Core Team, 2020) and the package ggplot, version 3.5.2, as well as MATLAB (2021, MathWork Inc.) and GraphPad Prisma.

**Figures (with legends) and Tables**

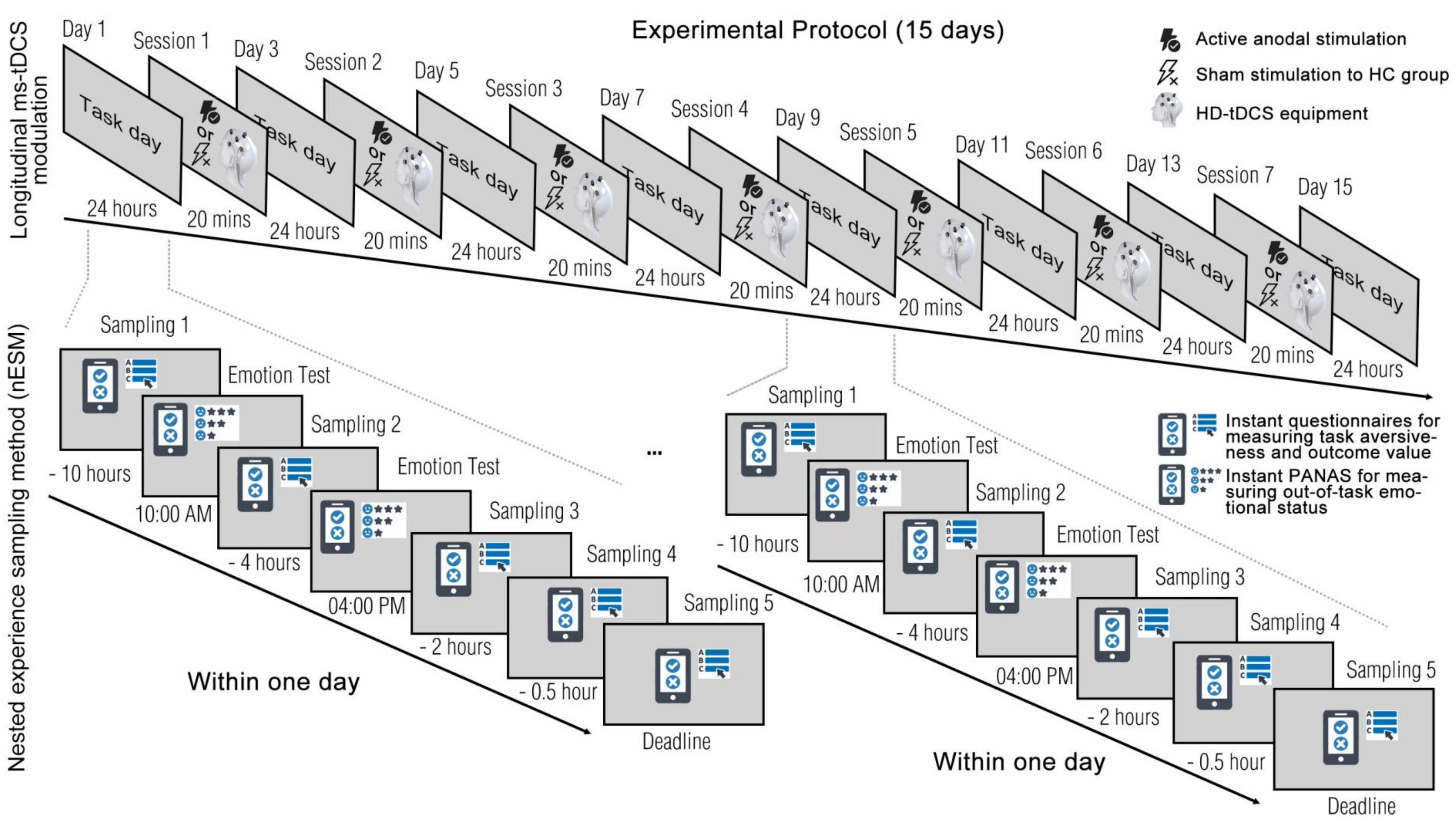

**Figure 1 Experimental diagram of this study.** The upper sub-graph illustrates the whole multiple sessions tDCS neuromodulation pipeline including seven sessions (days) and eight task-demanded days. Here, the “flash” icon indicates conducting tDCS neuromodulation (active anodal stimulation for active NM group and sham stimulation for sham NM group). This “+” label means a task-demanded day, where no stimulation is required for participants but all the covariates of interest should be measured by experience sampling methods. The bottom sub-graph reflects specific pipeline in task-demanded days. Participants were required to provide response at five progressive time moments nearing deadline for task aversiveness and outcome value in task-demanded days. In this diagram, the icon of “clock” symbolizes ecological momentary assessment for measuring instant task willingness and outcome value. In addition, twice tests for daily emotions (labeled by “+”) were added for participants at 10:00 and 16:00 as covariates of no interests to be adjusted.

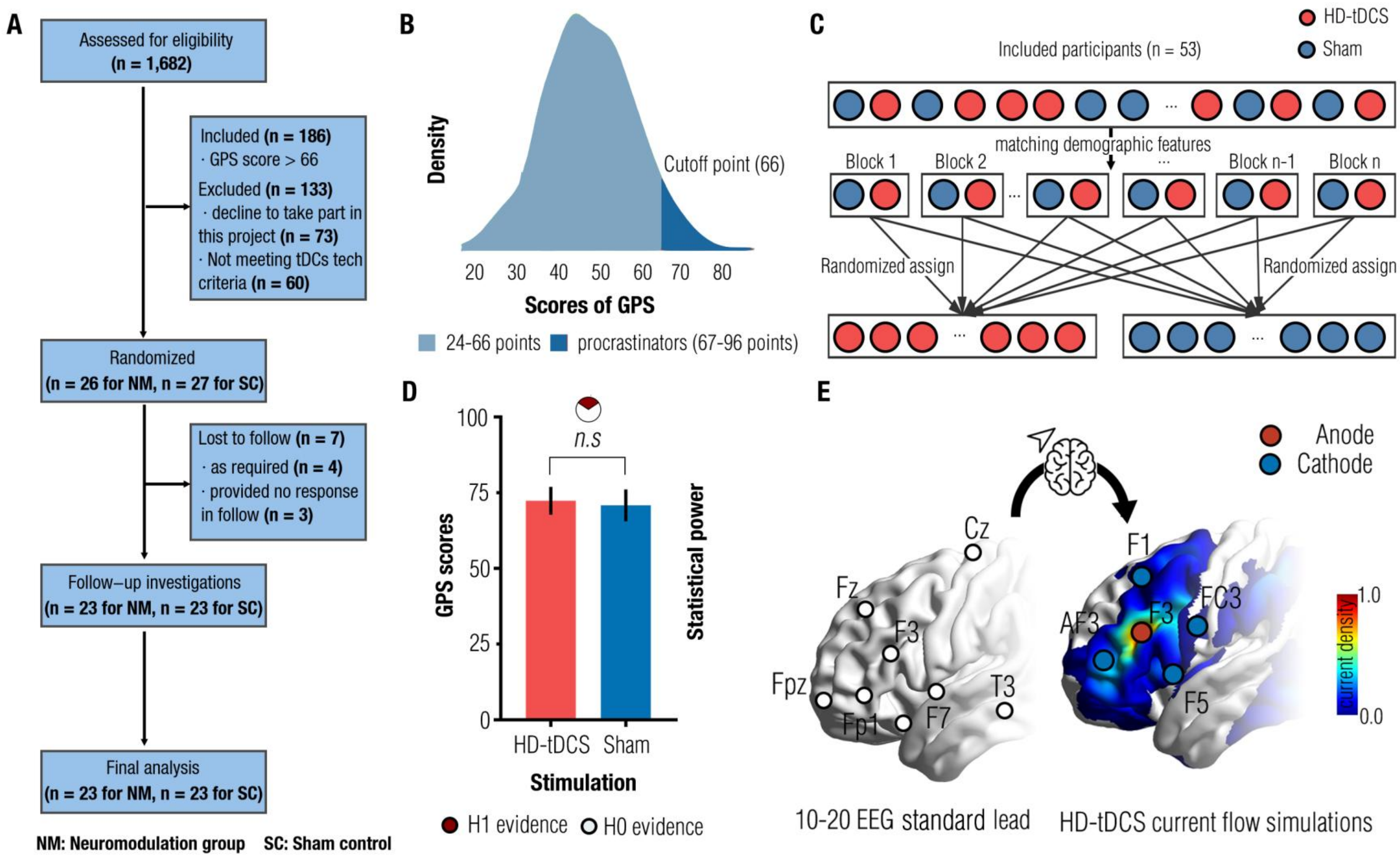
A
Assessed for eligibility
(n = 1,682)
Included (n = 186)
· GPS score > 66
Excluded (n = 133)
· decline to take part in this project (n = 73)
· Not meeting tDCs tech criteria (n = 60)
Randomized
(n = 26 for NM, n = 27 for SC)
Lost to follow (n = 7)
· as required (n = 4)
· provided no response in follow (n = 3)
Follow-up investigations
(n = 23 for NM, n = 23 for SC)
Final analysis
(n = 23 for NM, n = 23 for SC)
NM: Neuromodulation group SC: Sham control
B
Density
Cutoff point (66)
20 30 40 50 60 70 80
Scores of GPS
24-66 points
procrastinators (67-96 points)
C
HD-tDCS
Sham
Included participants (n = 53)
matching demographic features
Block 1
Block 2
Block n-1
Block n
Randomized assign
Randomized assign
D
100
75
50
25
0
n.s
GPS scores
Statistical power
HD-tDCS
Sham
Stimulation
H1 evidence
H0 evidence
E
Anode
Cathode
Cz
Fz
F3
Fpz
Fp1
F7
T3
F1
FC3
AF3
F3
F5
1.0
0.0
current density
10-20 EEG standard lead
HD-tDCS current flow simulations

**Figure 2** Flow diagram of CONSORT (A) and partial details of randomized groups (B-D) and neural locations of electric pole (E). B plots the distribution of all the participants procrastination scores (GPS = general procrastination scale). C detailed what the full randomized block design is. D shows the comparison between active neuromodulation group and sham control for procrastination scores. E indicates the pipeline to determine the location of electric pole. The 10-20 EEG standard lead is used to locate the left dorsolateral prefrontal cortex (DLPFC) initially, and the neuronavigation is further utilized to locate the exact location of this targeting region (i.e., left DLPFC).

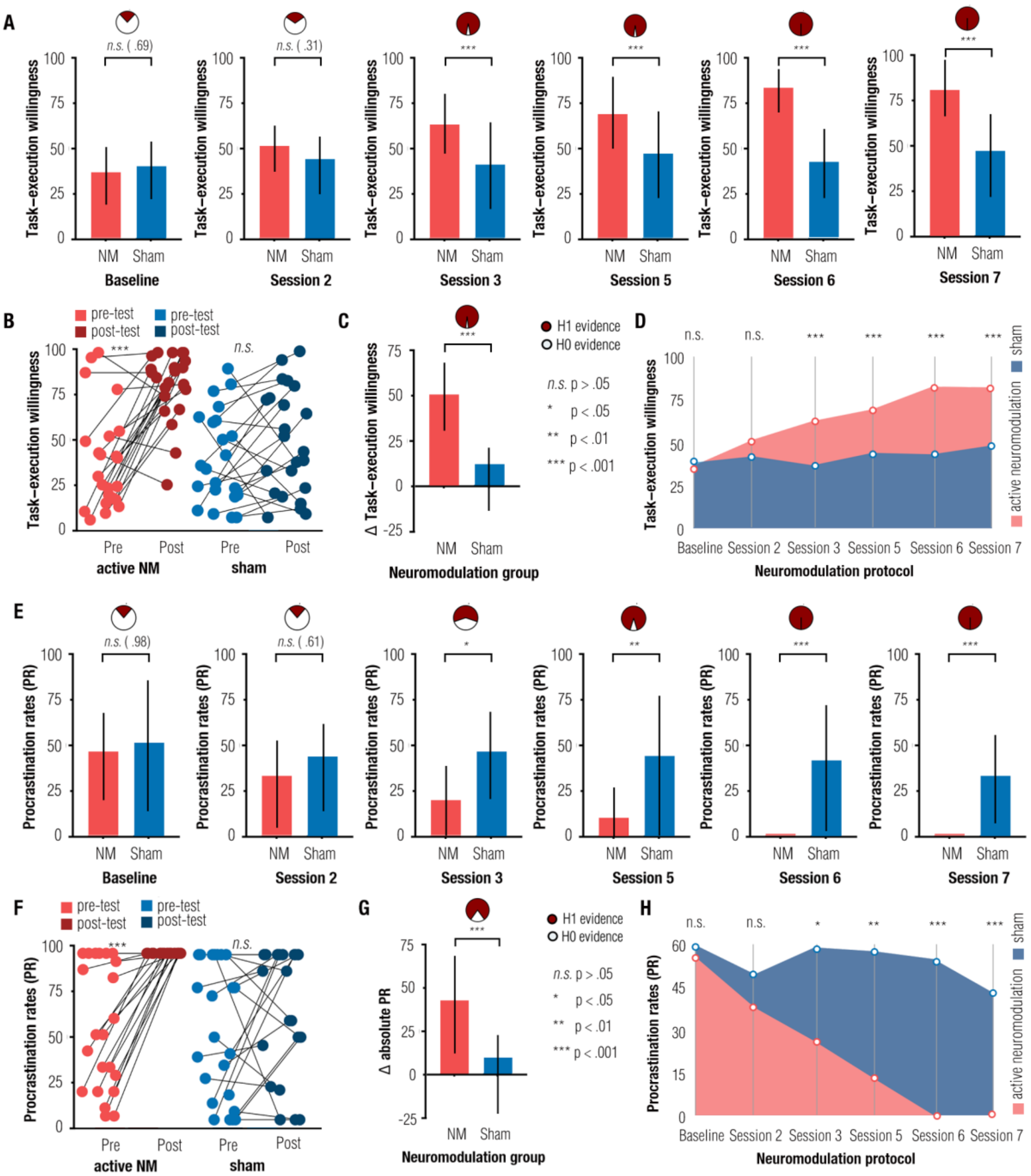

**Figure 3** Results of neuromodulation effects to task-execution willingness and procrastination rates (PR). A shows the effects of neuromodulation to increase task-execution willingness for both active group and sham control across sessions that included in formal analysis (session 0 (baseline), 2, 3, 5, 6, 7). B illustrates the effects of whole neuromodulation round to task-execution willingness for both groups. C plots the changes of task-execution willingness for both groups after neuromodulation. D provides a line chart to show the changes of task-execution willingness across each session that included in formal analysis. E shows the effects of neuromodulation to reduce PR for both active group and sham control across sessions that included in formal analysis. F illustrates the effects of whole neuromodulation round to PR for both groups. G plots the absolute changes of PR for both groups after neuromodulation. H provides a line chart to show the changes of PR across these sessions that included in formal analysis. Pie graph for each comparison represents the corresponding result of Bayesian factor inference, with brown piece for supporting H1 evidence and white piece for supporting H0 evidence. Each bar indicates mean value, and each line placed onto the bar reflects standard deviation (SD). To ensure transparency and rule out outlier-driven effects, individual data points for all participants (N=23 per group) are overlaid on the bars using a jittered distribution to prevent overplotting of identical values.

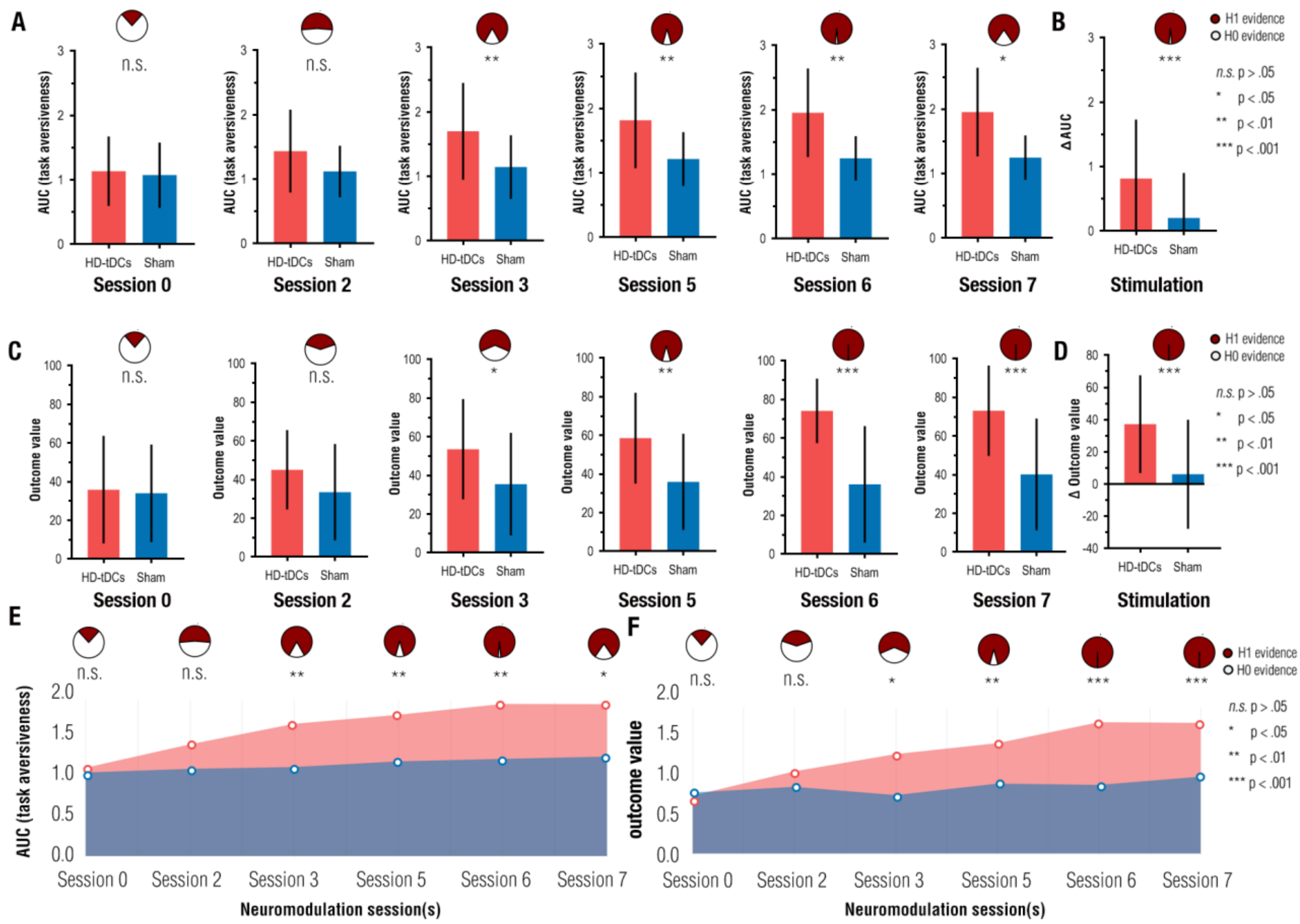

**Figure 4** Results of neuromodulation effects to task aversiveness and outcome value. A shows the effects of neuromodulation to increase AUC of task aversiveness for both active group and sham control across sessions that included in formal analysis (session 0 (baseline), 2, 3, 5, 6, 7). Higher AUC indicates lower task aversiveness as a given task is increasingly executed. B plots the changes of AUC of task aversiveness for both groups after neuromodulation. C shows the effects of neuromodulation to increase outcome value for both active group and sham control across sessions that included in formal analysis. D plots the changes of outcome value for both groups after neuromodulation. E provides a line chart to show the changes of AUC of task aversiveness across these sessions that included in formal analysis. F provides a line chart to show the changes of outcome value in the same manner. Pie graph for each comparison represents the corresponding result of Bayesian factor inference, with brown piece for supporting H1 evidence and white piece for supporting H0 evidence. Each bar indicates mean value, and each line placed onto the bar reflects standard deviation (SD).

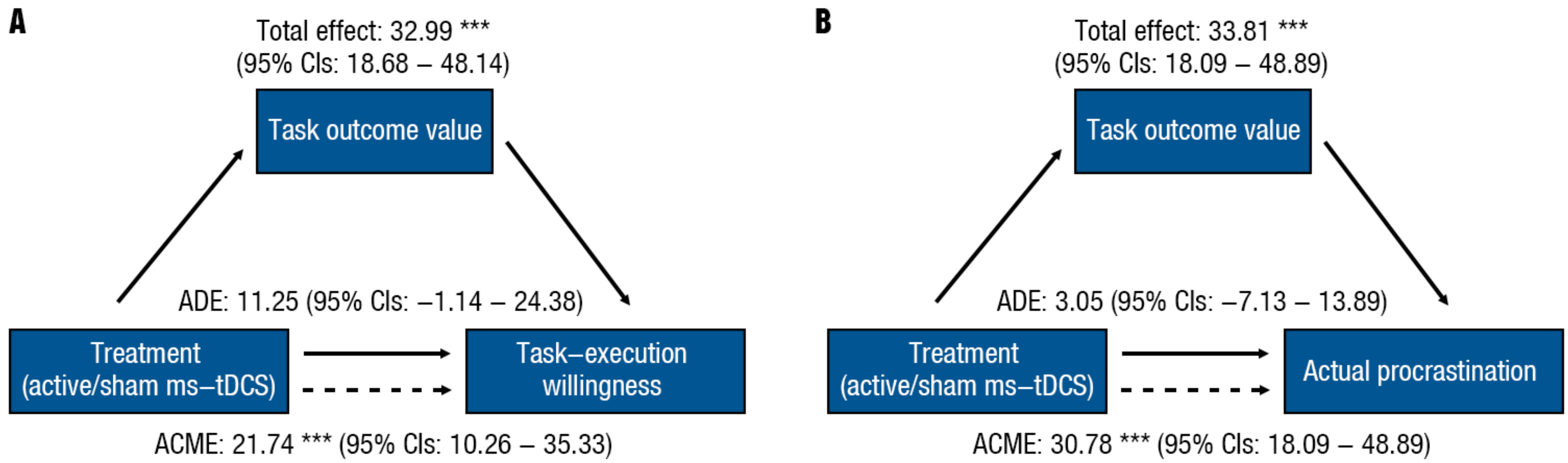

**Figure 5** Results of Quasi-Bayesian model for the mediated role of increased task outcome value in the association between neuromodulation and task-execution willingness (A) and actual procrastination rates (B). ADE = average direct effect, ACME = average causal mediation effect, CI = confidence interval.

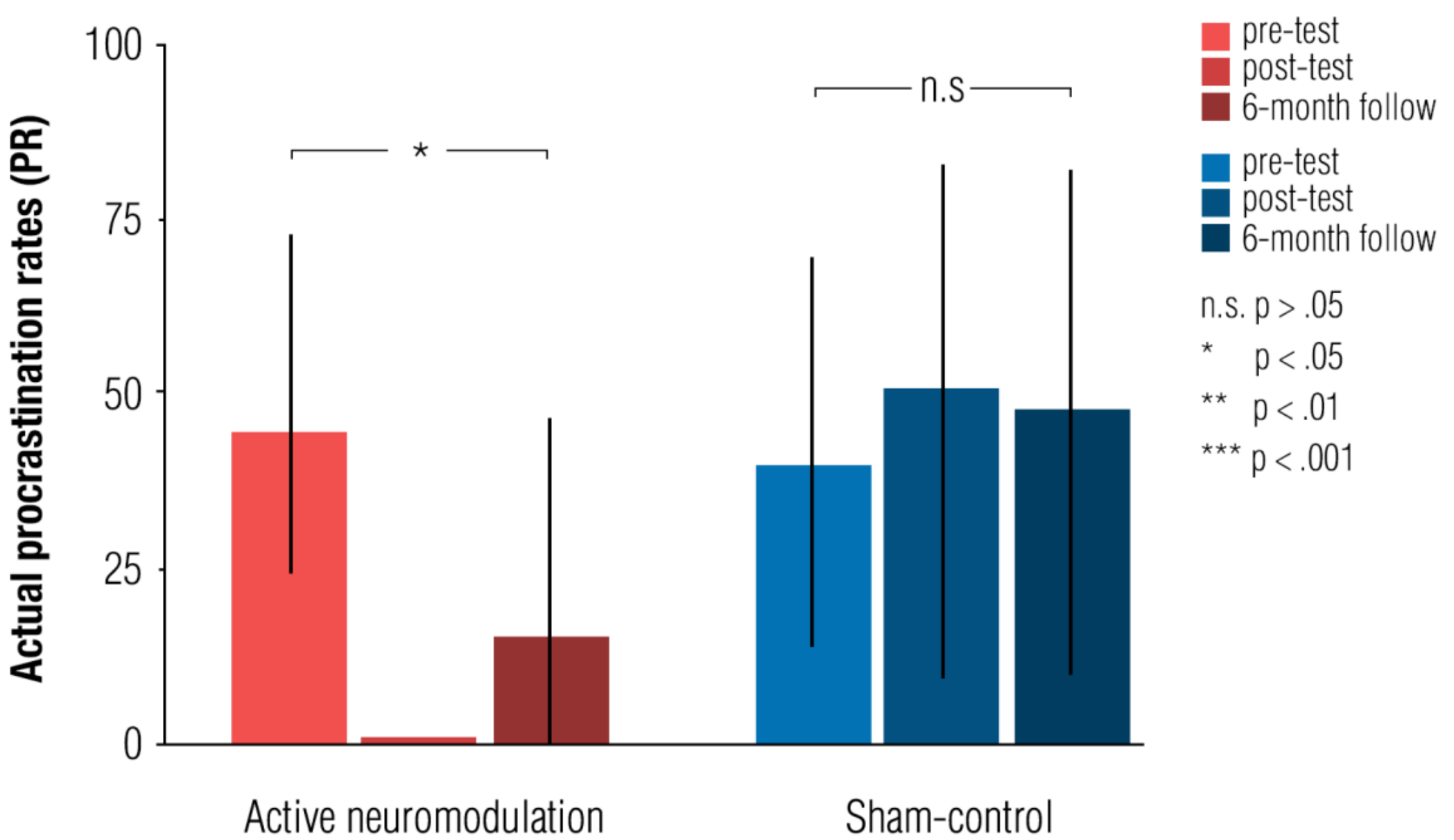

**Figure 6** Changes of actual procrastination rates among pre-test, post-test and 6-month follow-up for active neuromodulation group and sham-control group. Pre-test means to test the actual procrastination rates before first HD-tDCS neuromodulation. Post-test means to test the actual procrastination rates after last neuromodulation. The 6-month follow-up means to re-investigate the actual procrastination rates after last neuromodulation for 6 months.

| | active NM | | SC | | $P$-value ($BF_{10}$) |
|---|---|---|---|---|---|
| | Male | Female | Male | Female | |
| Gender | 3 | 20 | 3 | 20 | .99 ( - ) |
| Age | 19.61 ± 0.78 | | 22.22 ± 1.44 | | 0.08 (1.03) |
| SES | 2.17 ± 0.65 | | 2.34 ± 0.64 | | 0.38 (0.41) |
| Anxiety | 48.48 ± 6.52 | | 47.30 ± 6.87 | | 0.56 (0.33) |
| Depression | 47.13 ± 8.27 | | 47.50 ± 9.28 | | 0.88 (0.29) |
| Procrastination | 71.00 ± 5.47 | | 72.07 ± 4.77 | | 0.26 (0.48) |

**Table 1** Demographic information for included participants. NM represents active neuromodulation group and sham indicates sham-control group. Anxiety symptoms were measured by State-Trait Anxiety Inventory (STAI). Depression symptoms were tested by Self-Rating Depression Scale (SDS). BF10 describes the Bayesian evidence strength to support alternative hypothesis, with > 3 for a strong evidence.

| | ***β*** | **S.E** | ***P* value** | **Odd Ratio (OR)** | **adj $R^2$** |
|---|---|---|---|---|---|
| Δ task aversiveness | 0.62 | 0.42 | 0.13 | 1.85 | .29 |
| Δ outcome value | 0.85* | 0.42 | 0.04 | 2.34 | |

**Table 2** Summary for general linear model in predicting changes of task aversiveness and outcome value to actual procrastination. S.E. means standard error. * p < .05.